\documentclass[aps,pre,showkeys,
superscriptaddress,
longbibliography,
reprint,
twocolumn, 
nofootinbib,
floatfix]{revtex4-1}

\usepackage[utf8]{inputenc}
\usepackage[T1]{fontenc}
\usepackage{amsmath}
\usepackage{amsfonts}
\usepackage{graphicx}
\usepackage{tikz}
\usepackage{subcaption}
\usepackage[colorlinks=true,hyperfootnotes=true,breaklinks=true]{hyperref}
\usepackage{cleveref}

\DeclareMathOperator{\sign}{\mathrm{sign}}

\begin{document}

\title{Mean-field approximation for structural balance dynamics in heat-bath}

\author{Krzysztof Malarz}
\thanks{\includegraphics[scale=0.1]{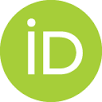}~\href{https://orcid.org/0000-0001-9980-0363}{0000-0001-9980-0363}}
\email{malarz@agh.edu.pl}
\affiliation{AGH University of Science and Technology,
Faculty of Physics and Applied Computer Science,\\
al. Mickiewicza 30, 30-059 Krak\'ow, Poland}
\author{Janusz A. Ho{\l}yst}
\thanks{\includegraphics[scale=0.1]{ORCID.png}~\href{https://orcid.org/0000-0003-2645-0037}{0000-0003-2645-0037}}
\email{janusz.holyst@pw.edu.pl}
\affiliation{Warsaw Technical University,
Faculty of Physics,\\
ul. Koszykowa 75, 00-662 Warszawa, Poland}

\begin{abstract}
A critical temperature for a complete signed graph of $N$ agents where time-dependent links polarization tends towards the Heider (structural) balance is found analytically using the heat-bath approach and the mean-field approximation as $T^c=(N-2)/a^c$, where $a^c\approx 1.71649$. The result is in perfect agreement with numerical simulations starting from the paradise state where all links are positively polarized as well as with the estimation of this temperature received earlier with much more sophisticated methods. When heating the system, one observes a discontinuous and irreversible phase transition at $T^c$ from a nearly balanced state when the mean link polarization is about $x_c=0.796388$ to a disordered and unbalanced state where the polarization vanishes. When the initial conditions for links polarization are random, then at low temperatures a balanced bipolar state of two mutually hostile cliques exists that decays towards the disorder and there is a discontinuous phase transition at a temperature $T^d$ that is lower than $T^c$. The system phase diagram corresponds to the so-called fold catastrophe when a stable solution of the mean-field equation collides with a separatrix, and as a result a hysteresis-like  loop is observed.
\end{abstract}
\date{\today}
\keywords{Heider balance; structural balance; signed networks; mean-field approximation; the first order phase transition; heat-bath algorithm}
\maketitle

\section{\label{sec:1:intro}Introduction}

Let $i$ and $j$ be persons and $k$ be an object that could be a third person, article, idea, event, etc. If the persons $i$ and $j$ possess the same attitude towards the object $k$ (e.g. they both like or both dislike it) then the theory of structural balance postulated by Heider \cite{Heider} says that it is more probable there is a positive relation between $i$ and $j$. 
On the other hand, if there is a disagreement between the attitudes of $i$ and $j$ towards $k$, then it is more likely that there is a negative relation between $i$ and $j$. In the case where $k$ is a person, the above propositions can be formed as the following rules: friend of my friend is my friend,  enemy of my enemy is my friend, friend of my enemy is my enemy and enemy of friend is my enemy. 

Structural balance theory met a lot of interest in social science and it was observed in many social groups when friendships and antipathies could be detected, see, e.g., References \onlinecite{Harary_1953,Cartwright_1956,Harary_1959,Davis_1967,Harary,Doreian_1996,Rambaran2015,Rawlings2017,Chiang2019, Toroslu_2022,Bramson_2022}. 
Antal, Krapivsky and Redner \cite{Antal_2005,Antal_2006} found a way to use the master equation for a description of possible dynamics of complex networks evolving toward the structural balance when mutual attitudes are described by binary variables of corresponding links.

Nowadays there are several attempts at theoretical description and computational simulation of the Heider balance, for example, when the attitudes are continuously changing variables \cite{Kulakowski2005,Marvel_2011,Gawronski_2015} or when attitudes are link attributes following from Hamming distances between nodes attributes \cite{Gorski_2020}. 
Often, an interpersonal relationship between two agents evolves, driven by the products of their relations (positive or negative) with their common neighbors.
In this way, two friendly or two hostile relations of two agents with their common neighbor improve their mutual relation, while their different relations with a neighbor drive their mutual relation to hostility.
For example, in References \onlinecite{Kulakowski2005,Marvel_2011}, the relations are represented by real numbers, and the system dynamics---by a set of differential equations. 
In these papers, the network of relations is a complete graph.
In References~\onlinecite{2005.11402,2008.06362} the relations are either positive or negative, and the dynamics is defined by a cellular automaton deterministic or with a thermal noise, and a local neighborhood of different range.
In References~\onlinecite{PhysRevE.104.024307,PhysRevE.105.054312} the relations are also discrete, the automaton rule is deterministic and the topology is a complete graph. In all these approaches, the target is a balanced state, i.e., a partition of the graph into two mutually hostile but internally friendly groups. 
When the sign of the relation of a pair as in Reference \onlinecite{PhysRevE.105.054312} is assumed to oscillate with the sum of products of relations over their neighbors, this target is reached immediately---in one time step---for each initial state \cite{Krawczyk_2022}.
For other approaches see References  \onlinecite{PhysRevE.90.042802,*Nishi_2014,*Kovchegov_2016,*Gorski_2017,*Krawczyk_2019,*PhysRevLett.125.078302,*Sheykhali_2020,*Pham_2021,*Deng_2021,*Arabzadeh_2021,*PhysRevE.105.054105,*Krawczyk_2015,*Stojkow_2020,*Krawczyk_2021}.

Recently \cite{PhysRevE.99.062302,PhysRevE.100.022303,2009.10136,2011.07501,2007.02128,2106.03054,2206.14226}, the Heider's dynamics has been enriched with social temperature $T$ \cite{social_temperature}. 
In Reference~\onlinecite{PhysRevE.99.062302} authors show that in the investigated system the first-order phase transition from ordered to disordered state is observed. A system of two coupled algebraic equations has been received using a mean-field approximation for an average link polarization and a correlation function between neighboring links polarization. The critical temperature of the complete graph consisting of $50$ nodes has been estimated by numerical solutions of these equations and agent based simulations confirmed the presence of the phase transition transition in such a model.                  
In this paper, we show a much simpler theoretical approach leading to the same conclusions. Similarly as in Reference~\onlinecite{PhysRevE.99.062302} we use the mean-field approximation but we find the critical temperature $T_c$ only from the average value of link polarization and we show that $T_c$ is proportional to number $M$ of different triangles containing a given link $T^c=M/1.71649\cdots$. If a network with all positive links is heated then the discontinuous phase transition takes place when the  mean link polarization decays to the critical value $x_c=0.796388$.
Our analytical results are well supported with computer simulation. 

\begin{figure}[htbp]
\begin{tikzpicture}[scale=1.1]
\draw[blue,very thick] (0,0) -- (1,0);
\node[below] at (0.5, 0) {$+$};
\node[below] at (0.5, 1.5) {(a)};
\draw[blue,very thick] (0,0) -- (0.5,0.866);
\node[left] at (0.25, 0.433) {$+$};
\draw[blue,very thick] (1,0) -- (0.5,0.866);
\node[right] at (0.75, 0.433) {$+$};
\draw[blue,very thick] (2,0) -- (3,0);
\node[below] at (2.5, 0) {$+$};
\node[below] at (2.5, 1.5) {(b)};
\draw[blue,very thick] (2,0) -- (2.5,0.866);
\node[left] at (2.25, 0.433) {$+$};
\draw[red,very thick, dashed]  (3,0) -- (2.5,0.866);
\node[right] at (2.75, 0.433) {$-$};
\draw[blue,very thick] (4,0) -- (5,0);
\node[below] at (4.5, 0) {$+$};
\node[below] at (4.5, 1.5) {(c)};
\draw[red,very thick, dashed]  (4,0) -- (4.5,0.866);
\node[left] at (4.25, 0.433) {$-$};
\draw[red,very thick, dashed]  (5,0) -- (4.5,0.866);
\node[right] at (4.75, 0.433) {$-$};
\draw[red,very thick, dashed]  (6,0) -- (7,0);
\node[below] at (6.5, 0) {$-$};
\node[below] at (6.5, 1.5) {(d)};
\draw[red,very thick, dashed]  (6,0) -- (6.5,0.866);
\node[left] at (6.25, 0.433) {$-$};
\draw[red,very thick, dashed]  (7,0) -- (6.5,0.866);
\node[right] at (6.75, 0.433) {$-$};
\end{tikzpicture}
\caption{\label{triangle}Heider's triads corresponding to balanced (first and third) and imbalanced (second and fourth) states}
\end{figure}

\section{\label{sec:1:Model}Model}
Consider a network of $N$ agents, and let us assume that the polarization of the links between two agents $i$ and $j$ is $x_{ij}=\pm 1$.
The dynamics towards the Heider balance can be written as 
\begin{equation}
x_{ij}(t+1)=\sign\left(\sum_k^{M_{ij}} x_{ik}(t)x_{kj}(t)\right),
\label{eq_disc}
\end{equation}
where the summation goes through $M_{ij}$ common nearest neighbors of the connected nodes $i$ and $j$, that is, $M_{ij}$ is the number of triangles that involve the link ${ij}$.
It means that for a single triangle system presented in \Cref{triangle} the first and third triangles are balanced in the Heider's sense (as friend of my friend is my friend and enemy of my friend is my enemy), while the second and the fourth are not.
In the latter case actors at triangles nodes either encounter the cognitive dissonance---as they cannot imagine how his/her friends can be enemies---or everybody hates everybody, which should lead to creation of a two-against-one coalition. Let us stress that formally the sum $\sum_k^{M_{ij}} x_{ik}(t)x_{kj}(t)$
can be treated as a local field acting on the link $x_{ij}$ and this field follows from the Hamiltonian \cite{Antal_2005,PhysRevE.99.062302}
\begin{equation}
\label{eq:Hamiltonian}
\mathcal{H}=-\sum_{i>j>k}^N x_{ij}x_{jk}x_{ki}. 
\end{equation}

\section{\label{sec:2:mfa}Mean-field analytical approach}

If one assumes that the link dynamics possesses a probabilistic character, then a natural form of updating rule \eqref{eq_disc} as in Heider dynamics can be:    
\begin{subequations}
\label{stoch1}
\begin{equation}
x_{ij}(t+1)=
        \begin{cases}
	+1 & \text{ with probability }p_{ij}(t),\\
	-1 & \text{ with probability }1-p_{ij}(t),
        \end{cases}
\end{equation}
where
\begin{equation}
    p_{ij}(t)=\frac{\exp(\xi_{ij}(t)/T)}{\exp(\xi_{ij}(t)/T)+\exp(-\xi_{ij}(t)/T)} \label{pMF}
\end{equation}
and
\begin{equation}
	\xi_{ij}(t)=\sum_{k}^{M_{ij}} x_{ik}(t)x_{kj}(t).
\end{equation}
\end{subequations}
Here, the positive variable $T$ can be considered as a social temperature
\cite{social_temperature}
(or a measure of the noise amplitude) and in the limit $T\to 0^+$ we have $p\to 1$ so \Cref{stoch1} reduces to \Cref{eq_disc}.
\Cref{stoch1} is nothing else but the heat-bath algorithm \cite[p.~505]{Binder_1997} for a stochastic version of \Cref{eq_disc} and thus it is ready for direct implementation in analytical investigations and computer simulations.

The expected value $\langle x_{ij}(t+1)\rangle$ in this approach is equal to
\begin{equation}
\label{eq_cont}
\langle x_{ij}(t+1)\rangle=\tanh\left( T^{-1} \sum_{k}^{M_{ij}} x_{ik}(t) x_{kj}(t) \right),
\end{equation}
where $\langle\cdots\rangle$ stands for a mean value related to the stochastic process defined by \Cref{stoch1}.  

The mean $\langle x_{ij}(t+1)\rangle$ is a continuous variable that can be negative or positive, and for $T\to 0^+$ the \Cref{eq_cont} reduces to \Cref{eq_disc}.

Now in our mean-field approximation we write the correlation function $\langle x_{ik}(t) x_{kj}(t)\rangle$ as the product $\langle x_{ik}(t)\rangle \langle x_{kj}(t)\rangle$. Let us note that such an approximation for the correlation function of link polarization is similar to the mean-field assumption used for the Ising model where correlations between spins $S_i$, $S_j$ are neglected, i.e. $\langle S_i S_j\rangle= \langle S_i\rangle \langle S_j\rangle$ see e.g. Reference~\onlinecite{PhysRev.149.380}. In fact, in Reference~\onlinecite{PhysRevE.99.062302} another mean-field approach was proposed where link-link correlations were considered but as we demonstrate in the Appendix \ref{app:Appendix} they can be disregarded in the thermodynamical limit of the studied system. Then instead of \Cref{eq_cont} we have
\begin{equation}
\label{eq_MF1}
\{\langle x_{ij}(t+1)\rangle\}\approx\tanh\left( \frac{1}{T} \sum_{k}^{M_{ij}} \{\langle x_{ik}(t)\rangle\} \{\langle x_{kj}(t)\rangle\} \right),
\end{equation}
where $\{\cdots\}$ denotes the average over all $N(N-1)/2$ available nodes' pairs. 

Now let us assume---also in agreement with the spirit of mean-field approximation---that all averages are the same
\begin{equation}
\label{eq:crude}
\{\langle x_{ij} \rangle\} = \{\langle x_{ik} \rangle\} = \{\langle x_{kj} \rangle \} = x.
\end{equation}
The above approximations are justified in the neighborhood of paradise\footnote{For paradise state all relations are friendly \cite{Antal_2005}.} state where the majority of triangles of type \ref{triangle}(a) are present.
However, as we show in further numerical simulations, the approach also works well in states far from paradise ($x\approx 0$).

It follows that we get    
\begin{equation}
x(t+1)=\tanh[ax^2(t)],
\label{eq_MF}
\end{equation}
where 
\begin{equation}
\label{eq:aMT}
a=M/T, 
\end{equation}
and $M=\{ M_{ij}\}$ is the average number of common neighbors of agents $i$ and $j$ (it also the number of different triads containing the edge $ij$). 

\begin{figure}
	\includegraphics[width=0.49\textwidth]{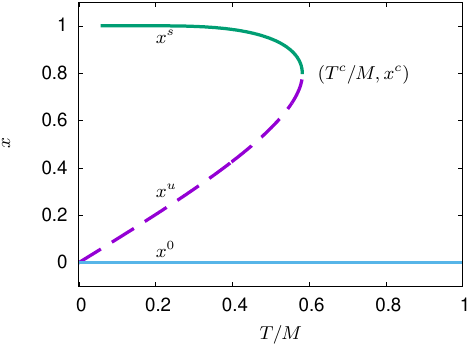}
	\caption{\label{fig:x_vs_a}Solutions for mean values of links polarizations in Heider balance resulting from the mean field approximation $x=\tanh(a x^2)$. At the point  $T^c/M=1/a^c\approx 0.582584\cdots$ ($x^c\approx 0.796388\cdots$) a discontinuous transition between the upper branch and the solution $x=0$ takes place.}
\end{figure}

We immediately recognize $x^0=0$ as a stable fixed point for any value of the $a$ parameter, and for $a\ll 1$ this is the only fixed point of \Cref{eq_MF}. However, for $a\gg 1$ there are two other fixed points, $x^u<x^s$ corresponding to unstable $x^u(a)$ and stable $x^s(a)$ solutions. In fact, $x^u$ 
is a separatrix between the domain of attractions of fixed points $x^0$ and $x^s$. When $a\gg 1$ then $x^u\approx 1/a$. When the parameter $a$ decreases from high values (this means that the temperature $T$ increases), then the fixed points $x^u$ and $x^s$ coincide together with the point $x^c$ for a certain value of $a=a^c$ (see \Cref{fig:x_vs_a}). 

This means that for $a>a_c$ the system is bi-stable and for $a<a^c$ the system is mono-stable. The above values $a^c, x^c$ can be received from a pair of transcendental algebraic relations that describe the fixed point and its tangency condition, namely
\begin{subequations}
\label{xc_ac}
\begin{equation}
	x^c=\tanh\left(a^c (x^c)^2\right)
\end{equation}
and
\begin{equation}
	2 a^cx^c  = \cosh^2\left(a^c (x^c)^2\right).
\end{equation}
\end{subequations}
The solutions (see \Cref{fig:x_vs_a}) are 
\begin{subequations}
\label{eq:xcacval}
\begin{equation}
\label{eq:xcval}
x^c_{\text{th}} \approx 0.796388\cdots
\end{equation}
and 
\begin{equation}
\label{eq:acval}
a^c_{\text{th}} =\tanh^{-1}(x^c_{\text{th}})/(x^c_{\text{th}})^2 \approx 1.71649\cdots.
\end{equation}
\end{subequations}

Let us note that since $x^c_{\text{th}}>0$ a system can express the phenomenon of hysteresis. It also means that we should not observe the values $0 <x <x^c_{\text{th}}$ as stable solutions.

\section{\label{sec:3:simul}Numerical estimation  of system critical temperature}

\begin{figure*}[htbp]
\includegraphics[width=.45\textwidth]{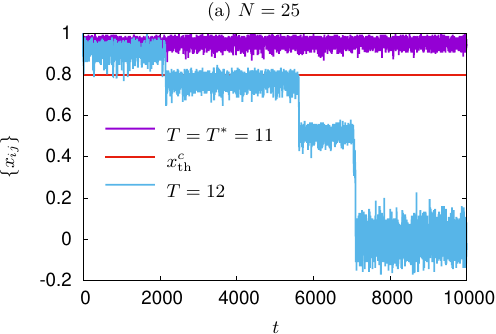}
\hfill\includegraphics[width=.45\textwidth]{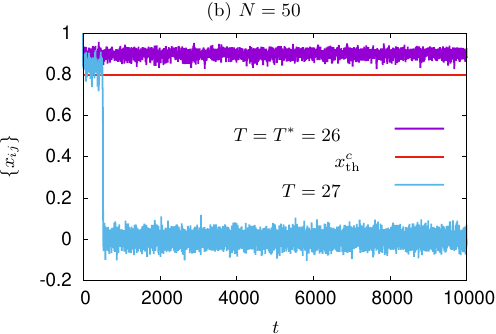}
\includegraphics[width=.45\textwidth]{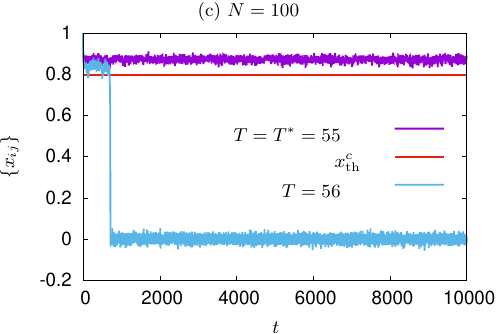}
\hfill\includegraphics[width=.45\textwidth]{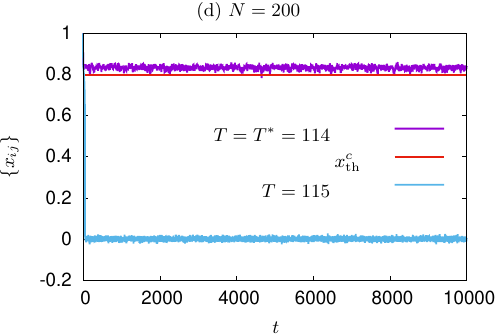}
\includegraphics[width=.45\textwidth]{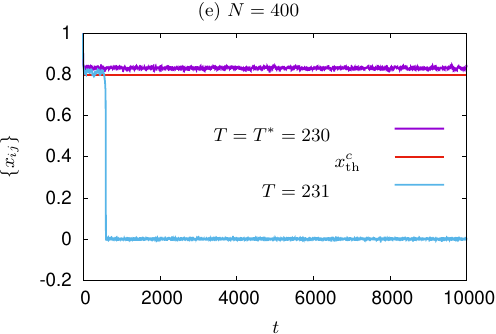}
\hfill\includegraphics[width=.45\textwidth]{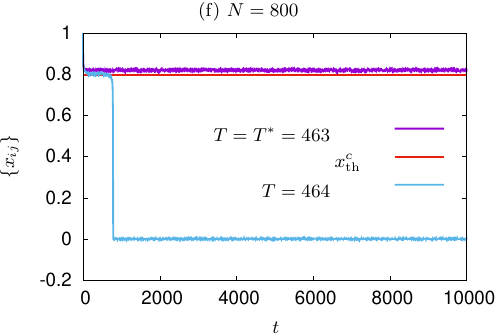}
\caption{\label{avexx_vs_time}The time evolution of the average values of all links polarization $\{ x_{ij}\}$ for various  social temperature $T$ and various system sizes $N$.
The starting point of simulation is homogeneous state  (paradise)  with $\{ x_{ij}\}=+1$ and the scanning temperature step is $\Delta T=1$.
When the temperature $T=T^*$ then averages $\{x_{ij}\}$ oscillate around values   that are close but always larger than the critical solution  $x^c$ (the solid red line given by the mean field approach \eqref{eq:xcval}) however when the temperature $T$ is slightly above $T^*$ the system evolves towards the state $\{x_{ij}\}\approx x^0=0$.
}
\end{figure*}

\begin{figure*}[htbp]
\includegraphics[width=.45\textwidth]{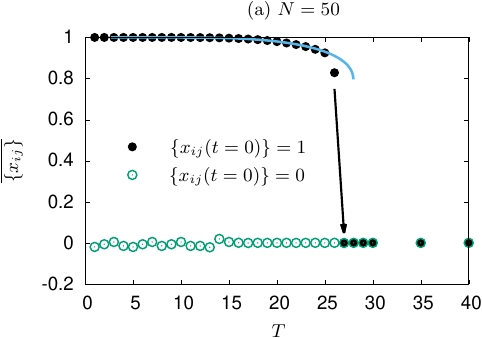}
\hfill\includegraphics[width=.45\textwidth]{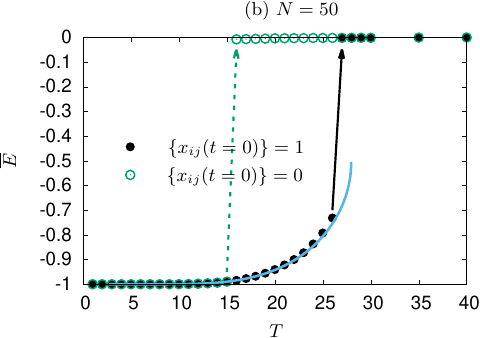}\\
\includegraphics[width=.45\textwidth]{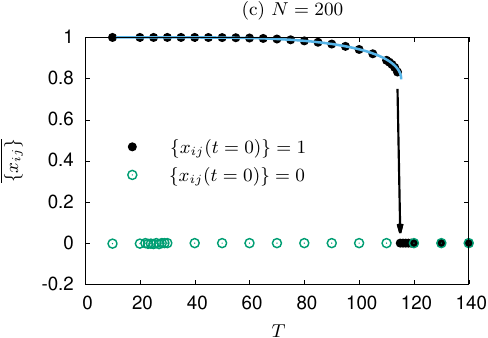}
\hfill\includegraphics[width=.45\textwidth]{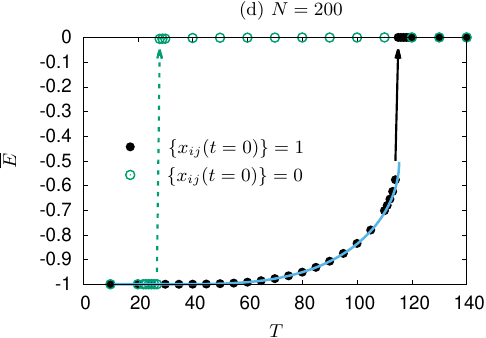}
\caption{\label{avexx_vs_T}(Color online). Panels (a, c) show the averages $\overline{\{ x_{ij}\}}$ and the panels (b, d) $\overline E$ as functions of the temperature  $T$ for complete graph with $N=50$ (a, b) and $N=200$ (c, d) nodes.
The averaging symbol $\overline{\cdots}$ stands for time average over the last $\tau=1000$ time steps of simulation.
The solid symbols correspond to the paradise state as  the starting point.
The open symbols stand for random initial state $\{ x_{ij}\}=0$.
The solid blue lines indicate the mean-field approximation predictions $\overline{\{x_{ij}\}}=x$ (a, c) and $\overline E =-x^3$ (b, d).
The solid black  or dashed green arrow marks the positions of the critical temperature $T^c$  or $T^d$ (see \Cref{sec:4:cliques}).
}
\end{figure*}

\begin{table}
\caption{\label{tab:}The numerically obtained values of $T^c_{\text{nu}}$ and $a^c_{\text{nu}}$ calculated basing on \Cref{eq:aC_TC} together with their estimated {expanded} uncertainties {$U(a^c_{\text{nu}})$}. The uncertainty of $T^c_{\text{nu}}$ is $u(T^c_{\text{nu}})=1/\sqrt{3}$, while $U(a^c_{\text{nu}})$ is calculated basing on \Cref{eq:u_aC}. Note that the differences between numerically estimated values and analytical results (values in the last column) are smaller than the uncertainties $U(a_{\text{nu}}^c)$.
}
\begin{ruledtabular}
\begin{tabular}{lllll}
	$N$ & $T^c_{\text{nu}}$ & $a^c_{\text{nu}}$ & {$U$}$(a^c_{\text{nu}})$ & $a^c_{\text{nu}}-a^c_{\text{th}}$ \\ \hline
	 25 &  11.5 & 2.00   & 0.30   & 0.28   \\ 
	 50 &  26.5 & 1.811  & 0.118  & 0.095  \\ 
	100 &  55.5 & 1.766  & 0.055  & 0.049  \\ 
	200 & 114.5 & 1.7293 & 0.0262 & 0.0128 \\ 
	400 & 320.5 & 1.7267 & 0.0130 & 0.0102 \\ 
	800 & 463.5 & 1.7217 & 0.0064 & 0.0052 \\ 
\end{tabular}
\end{ruledtabular}
\end{table}

To verify the analytical results in computer simulation, we directly apply \Cref{stoch1} to the time evolution of $x_{ij}$ for the complete graph with $N$ nodes. For the complete graph the average number of pair neighbors of nodes $ij$ is equal to $M=\{ M_{ij}\}=N-2$ and thus according to \Cref{eq:aMT} one should expect 
\begin{subequations}
\begin{equation}
\label{eq:aC_TC}
	a^c_{\text{nu}}=(N-2)/T^c_{\text{nu}}.
\end{equation}

To find the value of $T^c_{\text{nu}}$ we start the simulation with $T=0$ and scan the temperature $T$ with step $\Delta T$ and look for a value of $T^*$ for which $\langle x_{ij}\rangle$ is positive but for $T^*+\Delta T$ is zero. The true value of $T^c_{\text{nu}}$ is hidden somewhere in the interval $[T^*,T^*+\Delta T]$. We assume that $T^c$ value is uniformly distributed in the interval $[T^*,T^*+\Delta T]$ which allows us to estimate its uncertainty as $u(T^c_{\text{nu}})=\Delta T/\sqrt{3}$. The estimated value of $T^c_{\text{nu}}=(T^*+T^*+\Delta T)/2$. Based on \Cref{eq:aMT} we calculate the value of $a^c_{\text{nu}}$ and we can estimate its expanded uncertainty as
\begin{equation}
\label{eq:u_aC}
	U(a^c_{\text{nu}})={k}\left|\dfrac{\partial a}{\partial T}\right|_{T=T^c} u(T^c_{\text{nu}})={k}\dfrac{N-2}{(T^c_{\text{nu}})^2} u(T^c_{\text{nu}}),
\end{equation}
with the coverage factor $k=3$ \cite{nistTN1297}.
\end{subequations}

In \Cref{avexx_vs_time} the time evolution of $\{ x_{ij}\}$ for various values of social temperature $T$ and various system sizes $N$ is presented.
The starting point of the simulation is the homogeneous state (paradise) with $\{ x_{ij}\}=+1$ and the scanning temperature step is set to $\Delta T=1$.
The solid red line corresponds to $x^c_{\text{th}}$ given by \Cref{eq:xcval}.


The obtained critical temperatures $T^c_{\text{nu}}$ and their uncertainties $U(T^c_{\text{nu}})$ are collected in \Cref{tab:}.
The obtained values of $a^c_{\text{nu}}$ coincide nicely with those obtained analytically (see \Cref{eq:acval})
even under very crude assumptions given by \Cref{eq_MF}.
The values of $a^c_{\text{nu}}$ agree within {expanded} uncertainties {$U$}$(a^c_{\text{nu}})$ with its analytical partner $a^c_{th}$.

In \Cref{avexx_vs_T}(a), (c) the dependencies of $\overline{\{ x_{ij}\}}$ vs. $T$ for $N=200$ and $N=50$ are presented.
The averaging symbol $\overline{\cdots}$ represents the time average in the last $\tau=100$ time steps of the simulation, and
this time average should be approximately equal to the average $\langle \cdots \rangle$ used in \Cref{eq_MF1}, which comes from the ergodic theorem.
Solid symbols correspond to the starting point $\forall i,j: x_{ij}=+1$, while open symbols represent a random initial state $\{ x_{ij}\}=0$.
The latter recovers $x^0=0$ mentioned earlier.

In \Cref{avexx_vs_T}(b), (d) dependencies of the system energy density (average value of the Hamiltonian \eqref{eq:Hamiltonian} per triangle)  
\begin{equation}
E=-\dfrac{\sum_{i>j>k}x_{ij}x_{jk}x_{ki}}{\binom{N}{3}}
\end{equation}
are presented.
There is a discontinuous change in the mean system energy at the critical temperature, this corresponds to \Cref{avexx_vs_T}(b) in Reference~\onlinecite{PhysRevE.99.062302}. According to the mean-field approximation \Cref{eq:crude} we expect $E=-x^3$, and this approximation is marked by a solid blue line in \Cref{avexx_vs_T}(b), (d). Similarly to the numerically obtained values of $\overline{\{x_{ij}\}}$ also values of $\overline{E}$ agree fairly with the proposed mean-field approximation.

\section{\label{sec:4:cliques}Influence of mutually hostile  cliques on critical behavior}

In previous Sections we estimated analytically and numerically  the value of the system critical temperature $T_c$ when initial conditions were close to the paradise state.  We observed however that when a random initial state $\{x_{ij}(t=0)\}=0$ was used  in our numerical simulations   then a transition to a phase with a higher energy $\overline{E}=0$ took place at a temperature  $T^d$   that was lower than  $T^c$. Below we discuss the nature of  this transition (see also Reference~\onlinecite{2009.10136}).  

For random initial conditions the observed average $\{x_{ij}\}$ fluctuates around zero in time but the mean energy density is $\overline{E}=-1$ (see \Cref{avexx_vs_T}) in low temperatures.
In other words, this energy is the same as the system energy corresponding to the paradise state (with only positive links) $\{x_{ij}\}=1$. This means that the ground state of the system is degenerated \cite{Krawczyk_2017}. 
Although the fact that the state $\{ x_{ij}\}=0$ is stable in time follows from \Cref{eq_MF} its nature can be seen to be strange, since naively one would expect a picture of a disordered system corresponding to many unbalanced triangles with $\overline{E}=0$ and not $\overline{E}=-1$. 

\begin{figure}[htbp]
\begin{tikzpicture}[scale=1.8]
\draw[blue,very thick] (-1,-0.5) -- (-1,+0.5);
\draw[blue,very thick] (-1,-0.5) -- (-0.5,+1);
\draw[blue,very thick] (-1,-0.5) -- (-0.5,-1);
\draw[blue,very thick] (-1,+0.5) -- (-0.5,+1);
\draw[blue,very thick] (-1,+0.5) -- (-0.5,-1);
\draw[blue,very thick] (+1,-0.5) -- (+1,+0.5);
\draw[blue,very thick] (+1,-0.5) -- (+0.5,-1);
\draw[blue,very thick] (-0.5,-1) -- (-0.5,+1);
\draw[blue,very thick] (+0.5,-1) -- (+0.5,+1);
\draw[blue,very thick] (+0.5,-1) -- (+1,+0.5);
\draw[blue,very thick] (+0.5,+1) -- (+1,-0.5);
\draw[blue,very thick] (+0.5,+1) -- (+1,+0.5);
\draw[red,very thick, dashed] (-0.5,+1) -- (+0.5,+1);
\draw[red,very thick, dashed] (-0.5,-1) -- (+0.5,-1);
\draw[red,very thick, dashed] (-1,-0.5) -- (+1,-0.5);
\draw[red,very thick, dashed] (-1,+0.5) -- (+1,+0.5);
\draw[red,very thick, dashed] (-0.5,+1) -- (+0.5,-1);
\draw[red,very thick, dashed] (-0.5,-1) -- (+0.5,+1);
\draw[red,very thick, dashed] (-1,+0.5) -- (+1,-0.5);
\draw[red,very thick, dashed] (-1,-0.5) -- (+1,+0.5);
\draw[red,very thick, dashed] (-1,+0.5) -- (+0.5,-1);
\draw[red,very thick, dashed] (-1,-0.5) -- (+0.5,+1);
\draw[red,very thick, dashed] (-0.5,+1) -- (+1,-0.5);
\draw[red,very thick, dashed] (-0.5,-1) -- (+1,+0.5);
\draw[red,very thick, dashed] (-0.5,+1) -- (+1,+0.5);
\draw[red,very thick, dashed] (+0.5,+1) -- (-1,+0.5);
\draw[red,very thick, dashed] (-1,-0.5) -- (+0.5,-1);
\draw[red,very thick, dashed] (-0.5,-1) -- (+1,-0.5);
\end{tikzpicture}
\caption{\label{polarization}An example of bipolar  state with two mutually hostile  cliques of sizes  $h=4$. Each node possesses $(h-1)$ positive links to agents in the own clique and $h$ negative links to the second clique.}
\end{figure}

The truth is that the  stochastic evolution \Cref{stoch1}  starting from the unordered state $\{x_{ij}\}=0$ can lead to a polarized phase consisting of two paradises (cliques containing only friendship triangles \ref{triangle}(a)) of similar sizes (for entropic reasons) connected by hostile triangles \ref{triangle}(c)---see \Cref{polarization}. 
Such a balanced state composition has already been predicted for systems driven by the structural balance  by \citet{Cartwright_1956} in the middle `50. The phase was observed also in the model of constrained triad dynamics (CTD) introduced by Antal \cite{Antal_2005,Antal_2006} when a randomly selected link flips its temporal polarization $x_{ij}$ only if the total number of imbalanced triad decreases (see also Reference \onlinecite{2005.11402}). 
Such a case corresponds to \Cref{eq_disc} and it is equivalent to $T\to 0^+$ limit of heat-bath Equations \eqref{stoch1}.

Let us assume that the size of the entire group is an even number $N=2h$ and that each hostile group has the size $h$. Such a partition is the most probable division of the entire group \cite{Antal_2005,Antal_2006}  and it is then easy to show that the mean of $\{x_ {ij}\}$ is zero in such a bipolar phase. In fact, every node possesses $(h-1)$ positive links to agents in the own group and $h$ negative links to the second group. It follows that the mean link polarization $\{ x_{ij}\}$ is equal to $-1/(2h-1)$ and vanishes in the thermodynamic limit ($h\to\infty$).
Then, the number of triangles \ref{triangle}(a)
\begin{equation} 
n_\Delta(+3) = 2\binom{h}{3} = \frac{1}{3}h(h-1)(h-2) 
\end{equation}
and the number of triangles \ref{triangle}(c)
\begin{equation} 
n_\Delta(-1) = 2h\binom{h}{2} = h^2(h-1) 
\end{equation}
may be calculated with combinatorial analysis due to the  complete graph symmetry.
It follows that in the thermodynamic limit, there is a special ratio between the numbers of triangles \ref{triangle}(c) and number of triangles \ref{triangle}(a) in this phase
\begin{equation} 
\label{eq:3}
\lim_{h\to\infty} \frac{n_\Delta(-1)}{n_\Delta(+3)}=3
\end{equation} 
and due to the absence of unbalanced triangles \ref{triangle}(b) and \ref{triangle}(d) in this phase the frequencies of triads presented in \Cref{triangle} are $(\frac{1}{4},0,\frac{3}{4},0)$.

\begin{figure}[htbp]
\includegraphics[width=.23\textwidth]{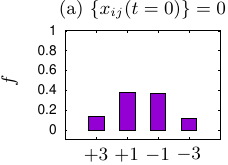}
\includegraphics[width=.23\textwidth]{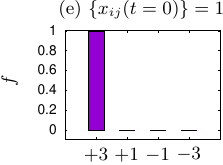}\\
\includegraphics[width=.23\textwidth]{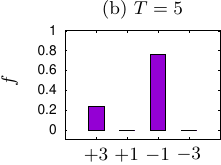}
\includegraphics[width=.23\textwidth]{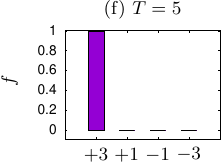}\\
\includegraphics[width=.23\textwidth]{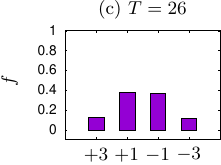}
\includegraphics[width=.23\textwidth]{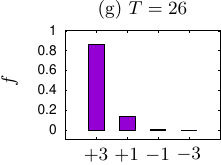}\\
\includegraphics[width=.23\textwidth]{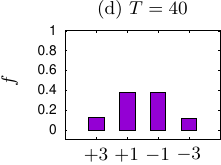}
\includegraphics[width=.23\textwidth]{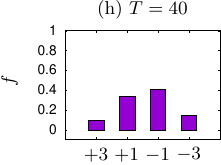}
\caption{\label{hist}Frequencies $f$ of appearing of triangles presented in \Cref{triangle} for $N=50$ and various initial states. In the first row the histograms for initial states are presented. In left (right) column histograms for random (paradise) initial states are presented. In rows 2--4 the histograms of frequencies of various triangles types in the final state of the systems are presented. The assumed temperatures $T$ are indicated in subfigures headlines.
    The numbers on abscissa axis are sum of polarizations $x_{ij}$ in a given triangle presented in \Cref{triangle}.}
\end{figure}

To verify our picture, we numerically investigated the frequencies $f$ of different triangles in the network when we start the evolution of the system from various initial conditions.
These frequencies are presented in \Cref{hist}.
When we start from $\{x_{ij}\}=0$ then for low temperatures (see \Cref{hist}(b)) the observed fraction $f(+3)$ of triangles \ref{triangle}(a) is three times lower than the fraction $f(-1)$ of triangles \ref{triangle}(c) as predicted by \Cref{eq:3}.
The fractions of triangles presented in \Cref{triangle} change abruptly from $(\frac{1}{4},0,\frac{3}{4},0)$ to $(\frac{1}{8},\frac{3}{8},\frac{3}{8},\frac{1}{8})$ at a critical temperature $T_d$ (see \Cref{hist}(c)--(d), cf. also Figure 10 in Reference~\onlinecite{2106.03054}).
The latter distribution of various types of triads is also kept in high temperature when we start the temporal system evolution from the paradise state (see \Cref{hist}(h)) and it corresponds to the probabilities of three, two, one, and zero successes in three Bernoulli trials when the probability of success is equal to $\frac{1}{2}$.

The above results mean that the polarized state with two hostile cliques exists only at low temperatures and disappears abruptly at $T=T^d$.
Let us stress that cliques emerge from random initial conditions in our numerical simulations.
A similar discontinuous transition was observed in CTD dynamics in References~\onlinecite{Antal_2005,Antal_2006} when the initial density of positive links and the average $\{x_{ij}(t=0)\}$ were changed in a continuous way.   
The analytical approach developed in Reference~\onlinecite{Antal_2006} shows that the critical value of $\{x_{ij}(t=0)\}$ for this transition should be equal to zero.
When this average is smaller, then the system reaches an equilibrium in the two cliques state, and above this threshold the system evolves towards the paradise.
Numerical simulations for the CTD model show a slightly higher value of this threshold \cite{Antal_2006}.
This discontinuous transition can be understood using our mean-field approach developed in \Cref{sec:2:mfa} as a result of the multistability of the system presented in \Cref{fig:x_vs_a}.
In fact, the unstable solution $x^u$ of \Cref{eq_MF} in \Cref{fig:x_vs_a} divides the set of initial conditions into two basins of attraction \cite{Ott_2002,*Kacperski_1996}.
When $\{ x_{ij}(t=0)\}$ is below $x^u$ the system evolves towards the solution $x^0=0$
and for an initial condition above this curve the system evolves towards $x^s$.
Since our model at $T=0$ corresponds to the CTD model and the separatrix $x^u$ in \Cref{fig:x_vs_a} approaches zero at this limit, the discontinuous disappearance of the polarized two-cliques phase observed in Reference~\onlinecite{Antal_2006} can be explained by crossing the separatrix curve. This critical behavior is equivalent to the bifurcation diagram corresponding to the so-called \emph{fold catastrophe} \cite{Scheffer_2009} and similar discontinuous phase transitions resulting from system bistability were observed in many systems, including pairs of weakly interacting Ising networks \cite{PhysRevE.80.031110,*PhysRevE.74.011122}, a majority model of a network with communities \cite{PhysRevE.75.030101}, as well as an activity-driven temporal bilayer echo-chamber system \cite{PhysRevE.105.024125}. 

\section{\label{S:Conclusion}Conclusions}

In this paper we present a simple analytical approach, which describes the first-order phase transition observed in thermalized Heider's balance systems on the complete graphs.
The proposed mean-field approximation predicts that the critical temperature of the system is equal to $T^c=(N-2)/a^c$, where $N$ is the number of graph nodes, and $a^c\approx 1.71649$. This temperature corresponds to a discontinuous transition from a critical state $\{x_{ij}\}= x^c\approx 0.796388$ (that is close to the paradise state) to an unbalanced state with the same number of positive and negative links $\{x_{ij}\}=x^0=0$. At the critical point $(T_c,x_c)$ the upper stable solution $x^s(T)$ for a fixed point of Equation \eqref{eq:aMT}---corresponding to a state of paradise dressed with thermal fluctuations at a given temperature $T$---coincides with the unstable branch $x^u(T)$ that is a separatrix, i.e. a boundary between initial conditions leading to the ``nearly paradise'' solution $x^s(T)$ or to the disordered solution $x^0$. For $T>T_c$ the solution $x^s(T)$ does not exist anymore and the system is always evolving towards $x^0$. This bifurcation scenario corresponds to the well-known fold catastrophe [see Figures \ref{avexx_vs_T}(a) and \ref{avexx_vs_T}(c)].

At the critical point, the system energy density changes from $\overline{E}=-(x^c)^3$ to the value $\overline{E}=0$ [see Figures \ref{avexx_vs_T}(b) and \ref{avexx_vs_T}(d)].
The results of computer simulations agree within the estimated uncertainties with our analytical calculations providing that initial conditions are close to the paradise state $\{x_{ij}(t=0)\}= 1$.

On the other hand, at any temperature $T$ when starting from a randomly selected state $x_{ij}=\pm 1$---when $\{x_{ij}(t=0)\}= 0$---we reach in the simulation only the solution $\{x_{ij}\}=0$ corresponding to the stable fixed point $x^0$. The solution can correspond to various patters of polarizations $x_{ij}$. At low temperatures, there is a phase consisting of two cliques of similar sizes that possess only positive internal links, but all inter-cliques links are negative. It follows that there are triangles consisting of all positive links or triads of one positive link and two negatives. It means that all the triads are balanced and that the energy density of the system is the same as in the paradise state, $\overline{E}=-1$, that is, degeneration of the ground is observed.
The signatures of such system division---i.e., the ratio $f(-1)\div f(+3)=75\%\div 25\%$---for low temperature noise level limit were also observed for diluted and densified triangulations (see Fig. 10 in Reference~\onlinecite{2106.03054}) and classical (Erd{\H o}s--R\'enyi) random graphs (see Fig. 7 in Reference~\onlinecite{2206.14226}).

At a certain temperature $T=T^d$ [for $\{x_{ij}(t=0)\}=0$] or $T=T^c>T^d$ [for $\{x_{ij}(t=0)\}=1$] another discontinuous phase transition occurs when the number of unbalanced triads [\ref{triangle}(b) and \ref{triangle}(d)] abruptly increases, and as a result the system energy density becomes zero [see dashed arrows connecting open symbols in Figures \ref{avexx_vs_T}(b) and \ref{avexx_vs_T}(d)]. This transition is not seen at a value of $\{x_{ij}\}$, which is zero below and zero above $T^d$ when $\{x_{ij}(t=0)\}=0$.

We stress that for $T^d<T<T^c$---depending on the initial conditions---the system energy is either close to the ground state or equal to zero [see Figures \ref{avexx_vs_T}(b) and \ref{avexx_vs_T}(d)]. Assuming initial conditions $\{x_{ij}(t=0)\}=1$ and heating the system from $T=0$ to $T=T^c$ decreases the mean value of $\{ x_{ij}\}$ along the curve $x^u(T)$ to the critical point $x^c$, but above $T^c$ we can only reach $\{ x_{ij}\}=0$.
The observed transition is irreversible, and cooling the system from $T>T^c$ towards $T\to 0^+$ will never reproduce positive values of $\{ x_{ij}\}>0$ and in this sense the hysteresis-like loop can be observed in the system, thus $(T_c,x_c)$ is a tipping point of our phase diagram \cite{Scheffer_2009}.

The critical temperature $T^c=26.2$ for the complete graph with $N=50$ nodes estimated in Reference~\onlinecite{PhysRevE.99.062302} is roughly in agreement with our estimate of $T^c$ for $N=50$. Finding analytically the value of a lower critical temperature $T^d$---where, starting with $\{x_{ij}(t=0)\}=0$, a special mixture of only balanced triangles \ref{triangle}(a) and \ref{triangle}(c) disappears---is beyond predictions of our approach, and it remains a challenge.

The most interesting result of these investigations for real social networks could be the observation that if such networks are driven by structural balance dynamics, then the balanced bipolar state seems to be the only possible state when the initial configuration of social links is completely random and the strengths of social interactions increase over time. To prohibit the emergence of such a polarized state, one could consider introducing additional attributes of interacting agents \cite{Gorski_2023}.

Finally, we note that Heider's theory may be applied for studies of international relations.
In Reference~\onlinecite{Antal_2005} authors presented the evolution of relations between countries as a prelude to World War I.
Further analysis of historical data is in progress \cite{Bartesaghi_2022}. 

\acknowledgments{We are grateful to Krzysztof Ku{\l}akowski, Krzysztof Suchecki and Piotr G\'orski for fruitful discussion and critical reading of the manuscript.
This research has received funding as RENOIR Project from the European Union’s Horizon 2020 research and innovation program under the Marie Sk{\l}odowska--Curie grant agreement No. 691152, by the Ministry of Science and Higher Education (Poland), grants Nos. W34/H2020/2016, 329025/PnH/2016. JAH was also supported by POB Research Centre Cybersecurity and Data Science of Warsaw University of Technology within the Excellence Initiative Program---Research University (IDUB) and by the Polish National Science Center, grant Alphorn No. 2019/01/Y/ST2/00058.}

\appendix
\section{\label{app:Appendix}}

In the mean-field approximation developed in Reference \onlinecite{PhysRevE.99.062302} the mean polarization of links is given as  
\begin{equation}
p=\langle x_{ij}\rangle=\tanh(\beta(N-2)q), \label{A1}
\end{equation}
where $\beta=1/T$ and $q=\langle x_{ik}x_{kj}\rangle$ is a correlation between links and 
\begin{equation}
q=\frac{e^{2(N-3)\beta q}- 2 e^{-2\beta p}+ e^{-2(N-3)\beta q}}{e^{2(N-3)\beta q}+ 2 e^{-2\beta p}+ e^{-2(N-3)\beta q}}.  \label{A2}
\end{equation}

Then the values $p$ and $q$ were found numerically as functions of the temperature $T$ from Equations~\eqref{A1} and \eqref{A2} in Reference~\onlinecite{PhysRevE.99.062302}. In our case, we receive the mean-field solution from the fix point of Equation~\eqref{eq_MF}, that is, $\langle x_{ij}\rangle=\tanh(\beta(N-2) \langle x_{ij}\rangle^2)$. However, one can easily find that when $N\gg 1$ then 
\begin{equation}
p^2\approx q. \label{A3}  
\end{equation}

In fact, in the thermodynamic limit, the difference between $q$ and $p^2$ is due to the second term of the nominator of Equation~\eqref{A2} which is $z=2\exp(-2\beta p)$ that should be equal to $2$ if Equation~\eqref{A3} is valid. However, one can write $z=2\exp(-(2p/T_c)(T_c/T))$ where $T_c$ is a critical temperature and since every link is influenced by many triangles in the Hamiltonian (\ref{eq:Hamiltonian}) thus $T_c\gg 1$. Thus, when the system is in the critical region and $T\approx T_c$ then since $p/T_c\ll 1$ thus $z\approx 2$ and $p\approx q^2$. The approximation also works well in the low temperature region $T\to 0^+$ since in such a case the terms $z$ and $2$ are much smaller than the first term in the nominator of Equation~\eqref{A2} and $q=1$.   

\end{document}